\documentclass[preprint2]{aastex}      
\usepackage{latexsym,times,flushrt}
\usepackage{graphicx}

\def\,{\relax\ifmmode\mskip\thinmuskip\else\thinspace\fi}
\catcode`\@=11
\def\.{.\spacefactor=\@m}

\shorttitle{Abundance of Atomic Sulfur at Io}
\shortauthors{Feaga et al.}

\begin{document}

\title{The Abundance of Atomic Sulfur in the Atmosphere of Io}

\author{Lori M. Feaga\altaffilmark{1}, Melissa A. McGrath\altaffilmark{2}, 
and Paul D. Feldman\altaffilmark{1}}

\altaffiltext{1}{Department of Physics and Astronomy, Johns Hopkins 
University, 3400 N. Charles Street, Baltimore, MD 21218; 
lanier@pha.jhu.edu, pdf@pha.jhu.edu}
\altaffiltext{2}{Space Telescope Science Institute, 3700 San Martin Drive, 
Baltimore, MD 21218; mcgrath@stsci.edu}

\begin{abstract}
	Observations with the Space Telescope Imaging Spectrograph aboard the 
Hubble Space Telescope have been used to constrain the atomic sulfur column 
density in Io's atmosphere.  The \ion{S}{1} $\lambda$1479 dipole allowed and 
forbidden transition multiplets have been resolved for the first time at Io, 
enabling the study of both optically thick and thin transitions from a single 
atomic species.  The allowed transitions contribute 62 $\pm$ 8\% and the 
forbidden transitions 38 $\pm$ 8\%, on average, to the total signal of the 
\ion{S}{1} $\lambda$1479 multiplets.  Using the optically thick and thin 
transitions of \ion{S}{1} $\lambda$1479 observed near the limbs of Io, we 
derive a tangential atmospheric sulfur column abundance of 
3.6$\times$10$^{12}$\,cm$^{-2}\,<\,{\cal N}_{s}\,<$\,\,1.7$\times$10$^{13}$\,cm$^{-2}$, 
which is independent of electron temperature and density.  A low density 
SO$_2$ atmosphere, ${\cal N}_{so_2}\sim$\,\,5-10$\times$10$^{15}$\,cm$^{-2}$, 
consistent with that inferred from other recent observations, is most 
consistent with these bounds.

\end{abstract}


\section{Introduction}

        Io's atmosphere plays a key role in populating the jovian 
magnetosphere with plasma and replenishing the plasma torus with sulfur and 
oxygen \citep{spe96}.  This tenuous atmosphere, first detected by the Voyager 
1 IRIS experiment over 20 years ago \citep{pea79}, has proven to be a 
challenge to characterize quantitatively.  It is composed primarily of SO$_2$ 
\citep{lel90,lel92} and SO \citep{lel96}.  The atomic by-products of SO$_2$, 
sulfur and oxygen, were first identified in the Io plasma torus 
\citep{bro81,dur83}, and more recently detected in Io's emission line 
spectrum \citep{bal87}.  Sodium \citep{bur98,gei99,bou00} and potassium 
\citep{bro01} are also present at much lower abundance than sulfur and 
oxygen.  The observed emissions are produced when the atmospheric sulfur 
and oxygen are collisionally excited by impinging electrons from the plasma 
torus.  The emissions are therefore diagnostic of the plasma interaction with 
Io's atmosphere, providing quantitative information about the electron 
temperature and density, in addition to the abundances of sulfur and oxygen.

	Despite the recent advances in quantifying Io's atmosphere, including 
investigating the inhomogeneity and impact of volcanic activity 
\citep{fel00,mcg00,spe00}, analyzing the airglow \citep{gei99}, and resolving 
the morphology \citep{roe99,ret00}, and the fact that several sets of sulfur 
and oxygen emission line data have been acquired and analyzed 
\citep{bal89,mcg00,oli01,wol01}, the abundances are still not well 
constrained.  Various observational measurements and theoretical models have 
placed loose bounds on the sulfur column density (${\cal N}_{s}$).  
Previously, electron densities and temperatures were assumed in order to 
estimate ${\cal N}_{s}$, resulting in limits of 
2.2$\times$10$^{12}$\,cm$^{-2}\,<\,{\cal N}_{s}\,<\,\,$7$\times$10$^{15}$\,cm$^{-2}$
\citep{bal87,mcg00}.  Recent observations make both optically thick and 
optically thin regimes available in a single exposure, allowing us to 
constrain ${\cal N}_{s}$ without assuming a specific $T_e$ or $n_e$.  We 
present such a temperature and density independent method for determining 
the column density in this paper.  

	Sulfur and oxygen abundances also provide a little-utilized 
diagnostic of the origin and nature of the SO$_2$ atmosphere.  Significant 
progress has been made in characterizing both the spatial distribution and 
temporal variability of the SO$_2$ atmosphere 
\citep{bal94,tra96,mcg00,fel00,spe00,str01}.  The current picture 
includes a higher column density equatorial, sub-solar region, 
${\cal N}_{so_2}\sim\,\,$1-5$\times$10$^{16}$\,cm$^{-2}$, with a strong 
poleward gradient, where ${\cal N}_{so_2}$ decreases to 
$<\,$10$^{15}$\,cm$^{-2}$ \citep{fel00,str01} and column densities 
$>\,$10$^{16}$\,cm$^{-2}$ above active volcanic regions \citep{mcg00,spe00}.  
It is obvious that both active volcanism and SO$_2$ frost sublimation are 
important atmospheric sources, although the relative importance of each 
mechanism has not been determined \citep{sum96,won00,mos01}.  The models 
show that sulfur can serve as a proxy for SO$_2$, with enhanced abundance 
relative to oxygen during and shortly after volcanic activity as compared 
to a sublimation atmosphere.  Sulfur displays similar trends as, and scales 
well with, the SO$_2$ abundance in most of the dayside models.  Determining 
and monitoring the sulfur abundance, can, in principle, provide a much-needed 
measure of the variability of Io's atmosphere.

	We describe the observations and explain the data reduction in 
Section 2.  We then give the motivation for and present the analysis of the 
opacity of the sulfur transitions and the atomic sulfur column density in 
Section 3.  Finally, we discuss the implications of our findings in Section 4.

\section{Observations and Data Reduction}

         The Space Telescope Imaging Spectrograph (STIS) aboard the 
Hubble Space Telescope (HST) has made it possible to study the sulfur 
emission near 1479\,\AA\ in detail.  The allowed multiplet, 
\ion{S}{1} $\lambda$1479 
(3s$^{2}$\,3p$^{3}$\,4s$'\,{}^{3}$D$^{0}\to\,$3s$^{2}$\,3p$^{4}\,{}^{3}$P), 
and the intercombination multiplet, \ion{S}{1}] $\lambda$1479 
(3s$^{2}$\,3p$^{3}$\,3d\,${}^{5}$D$^{0}\to\,$3s$^{2}$\,3p$^{4}\,{}^{3}$P), 
are both present in the data and are resolved for the first time.  The 
analysis that follows focuses on the separation of the 
\ion{S}{1} $\lambda$1479 multiplets and its implications.  Additional data 
collected with the International Ultraviolet Explorer (IUE), from 1986-1988, 
and the Faint Object Spectrograph (FOS) on HST, from 1992-1996, are 
re-examined in light of the STIS data.  We present only one representative 
spectrum per instrument in this paper.  We concentrate specifically on the 
following sulfur emission multiplets: the intercombination and allowed 
multiplets at \ion{S}{1} $\lambda$1479, the allowed multiplet 
\ion{S}{1} $\lambda$1814 
(3s$^{2}$\,3p$^{3}$\,4s$\,{}^{3}$S$^{0}\to\,$3s$^{2}$\,3p$^{4}\,{}^{3}$P), 
and the intercombination doublet \ion{S}{1}] $\lambda$1900 
(3s$^{2}$\,3p$^{3}$\,4s$\,{}^{5}$S$^{0}\to\,$3s$^{2}$\,3p$^{4}\,{}^{3}$P).  
For all but the 1479\,\AA\ multiplet, there are several years of data 
that confirm the trends and ratios found in our analysis.  Table 1 
summarizes the data analyzed, with the observations presented in this 
paper shown in bold.  The morphology of the STIS observations is identical 
to that described in \citet{ret00}, with resolved bright auroral emission 
regions near the Io equator at both sub- and anti-Jupiter locations.  
The equatorial spots are not resolved in the other observations presented 
here.
 
        The primary data set we analyze was acquired with STIS on HST using 
the 52$''\,\times\,$2$''$ slit and the G140M grating, which produces a 
monochromatic image of Io at each emission line detected in the spectral 
range 1440-1500\,\AA.  The set contains four $\sim$19 minute exposures taken 
on 1998 August 21-22 while Io was near western elongation, Io orbital 
longitude 285$\arcdeg$-305$\arcdeg$.  The coadded flux spectrum of the first 
two exposures is presented in Figure 1.  The data are acquired at a dispersion 
of 0.051\,\AA/pixel and spectral resolution of $\sim\,$0.5\,\AA, allowing 6 
blended components of the \ion{S}{1} $\lambda$1479 multiplets to be 
resolved.  The detected features correspond to both the electric dipole 
allowed and forbidden transitions shown in the Grotrian diagram in Figure 2.  
Atomic data for the transitions are given in Table 2.  The blended allowed 
transition wavelengths are 1474.1\,\AA, 1483.1\,\AA, and 1487.2\,\AA, while 
the blended forbidden transitions are at 1473.0\,\AA, 1481.7\,\AA, and 
1485.6\,\AA.
   
        These data were reduced in the following manner.  The rectified (x2d)
images were boxcar smoothed over a 3\,$\times\,$3 pixel region and 
background subtracted.  The wavelength dependent background level was 
determined by averaging the signal in a region spatially one Io diameter 
away from and comparable in size to the bright spectral features.  Next, 
several rectangular sub-images, centered on the brightest emission areas of 
the anti- and sub-jovian equatorial features, were created in order to isolate 
these spots and to ensure no overlap of emission.  Lastly, the data in the 
sub-images were compressed spatially into a one-dimensional array of flux vs. 
wavelength.  Because the spots are not located at the center of the aperture, 
their flux spectra have systematic wavelength offsets, which we have 
corrected for.  A Gaussian profile was fitted to each of the 6 resolved 
emission features to determine their relative strengths.  These Gaussians are 
overplotted on the data in Figure 3.

        We re-analyze a set of IUE observations of Io which span the spectral 
range 1150-1950\,\AA\ \citep{bal89}.  Low resolution $\sim$\,5\,\AA\ 
exposures of 13-14 hours in length were made over a three year period with 
the short-wavelength (SWP) camera.  We present a representative spectrum in 
Figure 4a which is centered on Io's western elongation on 1987 July 31; it 
was reduced following the same method as Ballester and smoothed using a 
5-element boxcar routine.  An additional step was taken to remove the 
reflected light continuum longward of 1700\,\AA\ by subtracting a solar 
spectrum scaled to the intensity of the underlying continuum.
   
        The last set of Io observations we re-analyze was made with the FOS.  
These observations cover 1600-2300\,\AA\ and were obtained over a period of 
four years using the G190H grating, which has a dispersion of 
1.45\,\AA/diode.  Exposures varied in length between 10 and 20 minutes.  The 
representative spectrum shown in Figure 4b was acquired while Io was 
emerging into sunlight from Jupiter's shadow on 1996 October 2.  In order to 
minimize scattered light contamination, only the first 7 groups of the data, 
corresponding to Io in eclipse, were used.  The data were reduced in the same 
manner as presented in \citet{mcg00} and were rebinned by 2 pixels and 
smoothed by 3.

	Several factors were considered in the analysis, including the 
influence of the emission morphology on the results, and possible 
contamination from SO$_2$ absorption.  The emission morphology (e.g., one or 
two spots visible) depends on Io's orbital geometry \citep{roe99,ret00}, 
which may affect the absolute magnitude of the measured fluxes.  When Io is 
at western elongation, as it is for part of the IUE and all of the STIS 
integration time, both equatorial spots are in view for at least part of the 
exposure.  For the FOS egress data, when Io is in shadow emerging from behind 
Jupiter, only one equatorial spot would be visible if resolved.  We assume 
that the opacity and the \ion{S}{1} $\lambda$1479 analysis we present are 
independent of the emission morphology.  Concerning contamination due to 
SO$_2$ absorption, since the SO$_2$ is more concentrated at lower altitude 
than the atomic sulfur in the atmosphere, the absorption should be minimal.  
Also, at the relevant wavelengths, the average SO$_2$ transmission is close 
to unity.  Therefore, the relative line strengths of the sulfur multiplets 
should not be significantly affected.

\section{Analysis}

\subsection{Allowed and Forbidden Components of \ion{S}{1} $\lambda$1479}

	It is desirable to separate the allowed from the forbidden 
transitions since the transition probabilities and the oscillator strengths 
for the two transitions are very different.  The forbidden transitions are 
much more likely to be optically thin than the allowed ones, meaning two 
regimes of opacity can potentially be sampled.  A quantitative determination 
of the relative contributions of the allowed and forbidden transitions can 
also be used to help analyze numerous data sets for which the multiplets are 
unresolved.

	A synthetic sulfur emission spectrum is created for the STIS data 
set by adding together 13 Gaussians of the same FWHM, one at each of the 
central wavelengths of the components of the \ion{S}{1} $\lambda$1479 
multiplet.  Each Gaussian is weighted by its optically thin relative 
strength as reported in \citet{mor91} and then the entire synthetic spectrum 
is scaled to the intensity level of the data.  A fit of the synthetic 
spectrum to the data is shown in Figure 3.  The flux from each line is 
determined by integrating the area under the Gaussian curves in Figure 3 and 
is reported in Table 3.  The allowed transitions contribute an average of 
62 $\pm$ 8\% and the forbidden transitions an average of 38 $\pm$ 8\% to the 
total signal of the \ion{S}{1} $\lambda$1479 emission.  Although the anti- 
and sub-jovian equatorial spots differ in flux by up to a factor of 2, 
the relative line ratios and the contribution of the allowed and forbidden 
components are consistent throughout the data set.

\subsection{Opacity of the Observed \ion{S}{1} Emissions}

	A synthetic sulfur emission spectrum is also created for the IUE 
and FOS data sets.  A unique FWHM is chosen for each data set to match the 
resolution of the instrument.  Again, Gaussians are placed at the central 
wavelengths for all lines of a given sulfur multiplet, weighted by its 
optically thin relative strength \citep{mor91}, and scaled to the data.  If 
the data match the synthetic spectrum, sulfur is considered optically thin 
in that emission; otherwise, it is considered optically thick.
 
	The fits for \ion{S}{1} $\lambda$1814 and \ion{S}{1}] $\lambda$1900
are overplotted on the data in Figure 4.  If optically thin, the transitions 
of the \ion{S}{1} $\lambda$1814 multiplet should be in a 5.2\,:\,3\,:\,1 
ratio and the \ion{S}{1}] $\lambda$1900 doublet will be in a 3.6\,:\,1 ratio.  
The ratios are calculated from their transition probabilities listed in 
Table 2.  We observe a ratio of 3.4\,:\,5.6\,:\,1 and 4.5\,:\,5\,:\,1 for the 
1814 multiplet using the IUE and FOS data respectively, and 3\,:\,1 and 
3.4\,:\,1 in the 1900 doublet.  Our results therefore indicate that the 
\ion{S}{1}] $\lambda$1900 doublet is optically thin and the 
\ion{S}{1} $\lambda$1814 multiplet is optically thick.  
Moreover, we have analyzed all the IUE and FOS data listed in Table 1 and 
find that in all existing observations of Io, the 1900\,\AA\ doublet is 
observed to be optically thin and the 1814\,\AA\ multiplet is observed to be 
optically thick.
 
        The allowed transitions of the \ion{S}{1} $\lambda$1479 multiplet 
have a theoretical ratio of 5.2\,:\,3.1\,:\,1 when optically thin 
\citep{mor91}.  We observe a ratio of 5.3\,:\,3.8\,:\,1 in the STIS data for 
the \ion{S}{1} $\lambda$1479 allowed multiplet.  Initially, our line ratios 
seemed to indicate that the \ion{S}{1} $\lambda$1479 allowed multiplet is 
optically thin since the blended line ratios are consistent with the optically 
thin values; however, this contradicts the fact that the 
\ion{S}{1} $\lambda$1814 multiplet, which has the same ground state and 
similar oscillator strength as \ion{S}{1} $\lambda$1479, is consistently 
optically thick in all existing observations of Io.  A more detailed analysis 
of the \ion{S}{1} $\lambda$1479 allowed transitions shows that because the 
lines are blends at STIS resolution, optical depth effects can be hidden by 
the blending.  The shorter wavelength transitions of a given blend, for 
example 1473.99 \AA, 1474.38 \AA, and 1483.04 \AA, are preferentially 
depleted at high optical depth.  The flux in these transitions is 
redistributed in the multiplet such that the fluxes in the longer wavelength 
transitions of the blend, 1474.57 \AA\, and 1483.23 \AA, are enhanced, and 
the 5\,:\,3\,:\,1 ratio of the blends is unaffected by this flux 
redistribution.

\subsection{Constraints on ${\cal N}_{s}$}	

	The optical depth of a transition is directly proportional to both 
the column density and the absorption cross section, 
${\tau} = {\cal N}{\sigma}$, with $\tau$ = 1 serving as the conventional 
boundary between the optically thin and thick regimes.  Therefore, generally 
speaking, an upper limit can be placed on the sulfur column density in the 
optically thin regime, and a lower limit in the optically thick regime.  If 
$\sigma$ is known, the sulfur column density can be constrained without 
assuming a specific $T_e$ or $n_e$.  
 
	The absorption cross section of photons in the core of an optically 
thin transition where only thermal Doppler broadening and radiative damping 
are considered is directly proportional to the oscillator strength, $f_{ij}$,
of the transition \citep{ost89}.  Assuming $\tau$ = 1, 
$${\cal N} = \frac{1}{\sigma} = \frac{m_ec\Delta\nu_D}{\sqrt{\pi}e^2f_{ij}} = 
\frac{m_ec}{\sqrt{\pi}e^2f_{ij}}\sqrt{\frac{2kT}{m_sc^2}}\nu_o.
\eqno{(1)}
$$
Several consistent values of the oscillator strength are published for the 
dipole allowed transitions \ion{S}{1} $\lambda$1814 and 
\ion{S}{1} $\lambda$1479.  Only one value of the oscillator strength exists 
for the majority of the forbidden transitions of \ion{S}{1}] $\lambda$1479.  
For the \ion{S}{1}] $\lambda$1900 transition, only two values of the 
oscillator strengths and transition probabilities exist (M\"uller 1968; 
Tayal 1998--see the discussion in \citet{mcg00} on the order of magnitude 
disagreement between these values).  The values and references for the 
transition probabilities and oscillator strengths we use are given in Table 2.

        Using equation (1) with $T$ = 1000K and the oscillator strength of 
$f_{ij}$ = 0.090 \citep{mor91} for the \ion{S}{1} $\lambda$1479 allowed 
multiplet, the lower bound for the sulfur column density is 
3.6$\times$10$^{12}$\,cm$^{-2}$.  A similar lower bound of 
2.8$\times$10$^{12}$\,cm$^{-2}$ is found using the $f_{ij}$ value of 0.093 
\citep{tay98} for the \ion{S}{1} $\lambda$1814 multiplet.  An upper bound of 
1.7$\times$10$^{13}$\,cm$^{-2}$ is found using the oscillator strength of the 
dominant \ion{S}{1}] $\lambda$1479 forbidden transition line, 1472.97 \AA, 
listed in \citet{mor91} as $f_{ij}$ = 0.019.  This result is lower than the 
previous upper bound for the atomic sulfur column density at Io, 
7$\times$10$^{15}$\,cm$^{-2}$, by more than 2 orders of magnitude 
\citep{mcg00}.  An explanation of this discrepancy is given in the discussion.
 
\subsection{Constraints on $T_e$ and $n_e$}

        Knowledge about the sulfur column density in turn allows us to place 
constraints on the electron temperature and density of the impacting plasma.  
The observed line of sight brightness is given by
$$
{B_{ij}} = \int { n_s n_e Q_{ij}(T_e) dl}
\eqno(2)
$$
where $n_s$ and $n_e$ are the sulfur and electron densities, respectively, and 
$Q_{ij}(T_e)$ is the temperature dependent electron excitation rate 
coefficient.  Although it has been shown by \citet{com98} and \cite{sau99} 
that the plasma flow and interaction around Io creates varying electron 
density and temperature profiles at Io, we assume that $n_e$ and $T_e$ are 
constant along the line of sight, invert equation (2), and arrive at an 
approximate relationship between the sulfur column density and the measured 
emission line brightness,
$$
{\cal N} \sim \frac{10^6\,B_{ij}}
{\overline{n}_e\,Q_{ij}(\overline{T}_e)}  \ \ \ (cm^{-2}) 
\eqno(3)
$$
where $B_{ij}$ is in units of Rayleighs, $\overline{n}_e$ is in units of 
cm$^{-3}$, and $Q_{ij}$ is in units of cm$^{3}$ s$^{-1}$.  For a particular 
transition from an upper state $j$ to a lower state $i$, the rate coefficient 
is \citep{ost89}
$$
{Q_{ij}(\overline{T}_e) = \frac{8.63\times10^{-6}\,{\overline{\Omega}_{ij}e^{-E_{ij}/kT_e}}}{\omega_i\,\sqrt{T_e}}},
\eqno(4)
$$
where ${\overline\Omega_{ij}}$ is the thermally averaged collision strength,
$\omega_i$ is the statistical weight of the lower state, $E_{ij}$ is the 
transition energy in units of eV, $k$ is the Boltzmann constant, and $T_e$ is 
the electron temperature in units of Kelvin.  These equations apply only to 
optically thin transitions, therefore, the \ion{S}{1}] $\lambda$1479 forbidden 
multiplet or the \ion{S}{1}] $\lambda$1900 intercombination doublet must be 
used in determining the electron temperature and density.  As there are no 
electron impact excitation cross-sections or collision strengths available for 
the \ion{S}{1}] $\lambda$1479 forbidden multiplet, the 
\ion{S}{1}] $\lambda$1900 intercombination doublet is used for the analysis.  
We use the electron excitation rate presented in Figure 9a of \citet{mcg00}.

	Combining the sulfur column density bounds from the 
\ion{S}{1} $\lambda$1479 analysis, the line brightnesses of the 
\ion{S}{1}] $\lambda$1900 intercombination doublet from the IUE data when Io 
was west of Jupiter, and equation (3), the density and temperature of the 
impacting electrons can be estimated.  For an electron temperature range of 
0.8-100 eV and several impacting electron density values between 500 and 
10000 cm$^{-3}$, the column density is plotted in Figure 5.  Both a low 
temperature solution, 1.5 eV $< T_e <$ 7 eV, and a high temperature 
solution, 7 eV $< T_e <$ 100 eV, are possible.  The canonical values of 5 eV 
and 2000 cm$^{-3}$ are consistent with the low temperature solution.  
Assuming the low temperature solution implies 
1000 cm$^{-3}\,<\,n_e\,<\,$5000 cm$^{-3}$.

\section{Discussion}
 
        The STIS observations presented here and used to calculate the atomic 
sulfur column density are tangential cuts through Io's atmosphere centered on 
the equatorial spots; therefore, the column densities are also tangential.  
\citet{wol01} has recently published a collection of four years of HST/STIS 
low resolution observations of near Io emissions and extended emissions in 
the 1150-1730 \AA\, wavelength range.  The photon fluxes measured at 1 Io 
radius for the \ion{S}{1} $\lambda$1479 multiplets varied between 
3$\times$10$^{-4}$\,cm$^{-2}$\,sec$^{-1}$ and 
3$\times$10$^{-3}$\,cm$^{-2}$\,sec$^{-1}$, corresponding to eclipse and 
western elongation, respectively.  These values are consistent with our 
HST/STIS photon fluxes for \ion{S}{1} $\lambda$1479, implying that the 
tangential sulfur column that \citet{wol01} were sampling should 
have values for ${\cal N}_{s}$ similar to ours.  \citet{mcg00} determined 
the vertical column density of S, SO, and SO$_2$ above specific locations on 
Io, including the Pele volcano.  With an estimated collision strength and 
electron impact excitation rate coefficient for the \ion{S}{1}] $\lambda$1900 
doublet, and assuming canonical plasma torus values of electron density and 
temperature, 1000 cm$^{-3}$ and 5 eV, they determine an ${\cal N}_{s}$ of 
1$\times$10$^{14}$\,cm$^{-2}$ at Pele, which is much larger than our upper 
bound of 1.3$\times$10$^{13}$\,cm$^{-2}$.  A column density that is 
significantly larger than our value should result in additional optical depth 
effects.  In Figure 6, the Pele flux spectrum published in Figure 2 of 
\citet{mcg00} is compared with the optically thin FOS synthetic spectrum 
described earlier, and confirms that the \ion{S}{1} $\lambda$1814 multiplet 
is more optically thick over Pele than in our IUE and FOS spectra, lending 
credence to the significantly higher value of ${\cal N}_{s}$ over Pele despite 
the uncertainty in the \ion{S}{1}] $\lambda$1900 electron excitation rate.  
This implies that ${\cal N}_{s}$ is larger in volcanic plumes than in the 
equatorial spots.  

	In the 1-D, steady state photochemical atmospheric models
presented by \citet{sum96}, both high and low density SO$_2$ atmosphere 
sub-solar profiles with various levels of vertical eddy mixing are 
calculated.  We have extracted the vertical column densities of several 
atmospheric components from their published profiles.  The ${\cal N}_{s}$ 
values corresponding to the high density SO$_2$ cases, 
${\cal N}_{so_2}$ = 10$^{18}\,$cm$^{-2}$, were much higher than the limits
from our analysis.  For example, with a low vertical eddy mixing 
coefficient of $k$ = 10$^{6}\,$cm$^{2}$\,sec$^{-1}$, ${\cal N}_{s}$ $\sim$ 
2$\times$10$^{17}\,$cm$^{-2}$; for a high vertical eddy mixing 
coefficient of $k$ = 10$^{9}\,$cm$^{2}$\,sec$^{-1}$, ${\cal N}_{s}
\sim\,\,$9$\times$10$^{14}\,$cm$^{-2}$.  On the other hand, for their low 
density case (${\cal N}_{so_2}$ = 8$\times$10$^{15}\,$cm$^{-2}$ with high 
eddy mixing $k$ = 10$^{9}\,$cm$^{2}$\,sec$^{-1}$) ${\cal N}_{s}$ $\sim$ 
6$\times$10$^{12}\,$cm$^{-2}$.  The corresponding tangential column density, 
assuming spherical symmetry and using the atmospheric sulfur number density 
profile in \citet{sum96}, is on the order of 4$\times$10$^{13}\,$cm$^{-2}$.  
This value is much more consistent with our results and the general overall 
picture of the SO$_2$ atmosphere described in the introduction.  In a 3-D high 
density SO$_2$ model atmosphere presented by \citet{won00}, they assume an 
initial sub-solar ${\cal N}_{so_2}$ of 6$\times\,$10$^{17}\,$cm$^{-2}$, which 
results in a tangential sulfur column density at the terminator, 
${\cal N}_{s}\sim\,\,$2$\times$10$^{15}\,$cm$^{-2}$, inconsistent 
with our upper limit to ${\cal N}_{s}$.  In the model in which the input 
SO$_2$ column is reduced to a lower density value of 
7$\times\,$10$^{16}\,$cm$^{-2}$, the tangential ${\cal N}_{s}$ is more 
consistent with our results at a value of 6$\times\,$10$^{13}\,$cm$^{-2}$.

	In order to predict what the atmospheric driving mechanism is, 
\citet{mos01} have modeled both an SO$_2$ frost sublimation atmosphere and a
Pele-type volcanically driven atmosphere.  Of the two models, our 
${\cal N}_{s}$ value is more consistent with the SO$_2$ frost sublimation 
atmosphere simply because this model has a lower sulfur column density.  
However, the true test to the Moses et al. (2001) models is to compare the 
sulfur to oxygen column abundances extracted from a single exposure.  When 
the sulfur is more abundant than the oxygen, the models imply a volcanically 
driven atmosphere.  None of the medium resolution data presented here had a 
simultaneous oxygen detection, so the ratio is not currently available, 
although this sulfur to oxygen relationship will be analyzed in future work 
with a more extensive data set.

\acknowledgments

The authors would like to thank Darrell Strobel and Warren Moos for insightful 
discussions on the atomic physics of the UV sulfur transitions.  We are 
also grateful to Brian Wolven and Kurt Retherford for help with the 
STIS data reduction.  Support for this work was provided by NASA through 
grant number HST-AR-09211.01-A from the Space Telescope Science Institute, 
which is operated by the Association of Universities for Research in 
Astronomy, Incorporated, under NASA contract NAS5-26555.

\clearpage

\begin{deluxetable}{lcccr}
\tabletypesize{\footnotesize}
\tablecaption{Summary of Io Observations \label{tbl-1}}
\tablewidth{0pt}
\tablecolumns{5}
\tablehead{
\colhead{Observation} & \colhead{UT Date} & \colhead{UT Time} &
\colhead{Io Diameter} & \colhead{Exposure Time} \\
\colhead{} & \colhead{} & \colhead{(hh:mm(:ss))} &
\colhead{($''$)} & \colhead{(s)} 
}\startdata
\multicolumn{5}{c}{IUE}  \\
\tableline
swp28708 & 1986 Jul 18 & 21:49 & 1.14 & 48600  \\
swp29430 & 1986 Oct 12 & 15:29 & 1.21 & 49800  \\
{\bf swp31440} & {\bf 1987 Jul 31} & {\bf 18:53} & {\bf 1.06} & {\bf 50400}  \\ 
swp31447 & 1987 Aug 1 & 16:57 & 1.07 & 48000  \\
swp34341 & 1988 Sep 27 & 17:02 & 1.12 & 48600  \\
swp34343 & 1988 Sep 28 & 14:49 & 1.12 & 47700  \\
\cutinhead{HST/FOS}
y0w00203-4 & 1992 Mar 22 & 13:18:10 & 1.11 & 1680  \\
y0w05402 & 1992 May 16 & 10:17:41 & 0.97 & 1245  \\
y1a10801 & 1993 Aug 4 & 02:44:55 & 0.84 & 851  \\
y2wta201 & 1995 Oct 9 & 12:48:19 & 0.86 & 738  \\
y3cwb101-5 & 1996 Aug 1 & 07:20:06 & 1.17 & 4920  \\
{\bf y3eta201} & {\bf 1996 Oct 2} & {\bf 19:24:22} & {\bf 0.98} & {\bf 800}  \\
\cutinhead{HST/STIS}
{\bf o4xm03050} & {\bf 1998 Aug 21} & {\bf 22:07:17} & {\bf 1.23} & {\bf 1155}  \\
{\bf o4xm03060} & {\bf 1998 Aug 21} & {\bf 22:30:02} & {\bf 1.23} & {\bf 1155}  \\
o4xm03070 & 1998 Aug 21 & 23:42:42 & 1.23 & 1145  \\
o4xm03080 & 1998 Aug 22 & 00:05:17 & 1.23 & 1145  \\
\enddata
\end{deluxetable}

\clearpage

\begin{deluxetable}{lcclccl}
\tabletypesize{\footnotesize}
\tablecaption{Atomic Data for SI \label{tbl-2}}
\tablewidth{0pt}
\tablecolumns{7}
\tablehead{
\colhead{Multiplet} & \colhead{$J_u$} & \colhead{$J_l$} & 
\colhead{Actual $\lambda$} & \colhead{Transition Probability} & 
\colhead{Oscillator Strength} & \colhead{Reference} \\
\colhead{} & \colhead{} & \colhead{} & \colhead{(\AA)} & \colhead{(s$^{-1}$)} & \colhead{} & \colhead{}
}\startdata

{\bf ${}^{5}$D$^{0}\to\,{}^{3}$P} & \nodata & \nodata & \nodata & \nodata & \nodata & \nodata \\
\nodata & 3 & 2 & 1472.97 & 4.20 $\times$ 10$^{7}$ & 1.91 $\times$ 10$^{-2}$ &  1 \\
\nodata & 2 & 2 & 1473.02 & 6.15 $\times$ 10$^{5}$ & 2.00 $\times$ 10$^{-4}$ &  2 \\
\nodata & 1 & 2 & 1473.07 & 3.91 $\times$ 10$^{4}$ & 7.63 $\times$ 10$^{-6}$ &  2 \\ 

\nodata & 2 & 1 & 1481.66 & 1.70 $\times$ 10$^{7}$ & 9.33 $\times$ 10$^{-3}$ &  1 \\
\nodata & 1 & 1 & 1481.71 & 4.28 $\times$ 10$^{5}$ & 1.41 $\times$ 10$^{-4}$ &  2 \\
\nodata & 0 & 1 & 1481.74 & 3.84 $\times$ 10$^{3}$ & 4.21 $\times$ 10$^{-7}$ &  2 \\

\nodata & 1 & 0 &1485.62 & 2.30 $\times$ 10$^{6}$ & 2.28 $\times$ 10$^{-3}$ &  1 \\

\tableline

{\bf ${}^{3}$D$^{0}\to\,{}^{3}$P} & \nodata & \nodata & {\bf 1478.50} & {\bf 1.65 $\times$ 10$^{8}$} & 
{\bf 9.01 $\times$ 10$^{-2}$} & {\bf 1} \\

\nodata & 3 & 2 & 1473.99 & 1.60 $\times$ 10$^{8}$ & 7.30 $\times$ 10$^{-2}$ &  1 \\
\nodata & 2 & 2 & 1474.38 & 5.00 $\times$ 10$^{7}$ & 1.63 $\times$ 10$^{-2}$ &  1 \\
\nodata & 1 & 2 & 1474.57 & 6.20 $\times$ 10$^{6}$ & 1.21 $\times$ 10$^{-3}$ &  1 \\

\nodata & 2 & 1 & 1483.04 & 1.20 $\times$ 10$^{8}$ & 6.60 $\times$ 10$^{-2}$ &  1 \\
\nodata & 1 & 1 & 1483.23 & 7.50 $\times$ 10$^{7}$ & 2.47 $\times$ 10$^{-2}$ &  1 \\

\nodata & 1 & 0 & 1487.15 & 8.70 $\times$ 10$^{7}$ & 8.65 $\times$ 10$^{-2}$ &  1 \\

\tableline

{\bf ${}^{3}$S$^{0}\to\,{}^{3}$P} & \nodata & \nodata & {\bf 1813.73} & {\bf 5.67 $\times$ 10$^{8}$} & 
{\bf 9.33 $\times$ 10$^{-2}$} & {\bf 2} \\
\nodata & 1 & 2 & 1807.31 & 3.17 $\times$ 10$^{8}$ & 9.32 $\times$ 10$^{-2}$ &  2 \\
\nodata & 1 & 1 & 1820.34 & 1.88 $\times$ 10$^{8}$ & 9.33 $\times$ 10$^{-2}$ &  2 \\
\nodata & 1 & 0 & 1826.24 & 6.23 $\times$ 10$^{7}$ & 9.34 $\times$ 10$^{-2}$ &  2 \\

\tableline

{\bf ${}^{5}$S$^{0}\to\,{}^{3}$P} & \nodata & \nodata & \nodata & \nodata & \nodata & \nodata \\
\nodata & 2 & 2 & 1900.29 & 6.60 $\times$ 10$^{4}$ & 3.60 $\times$ 10$^{-5}$ &  3 \\
\nodata & 2 & 1 & 1914.70 & 1.80 $\times$ 10$^{4}$ & 1.70 $\times$ 10$^{-5}$ &  3 \\

\enddata
\tablerefs{
(1) Morton 1991; (2) Tayal 1998; (3) M\"{u}ller 1968.}
\end{deluxetable}

\clearpage

\begin{deluxetable}{cclll}
\tabletypesize{\footnotesize}
\tablecaption{Measured Fluxes for SI \label{tbl-3}}
\tablewidth{0pt}
\tablecolumns{5}
\tablehead{
\colhead{Multiplet} & \colhead{Central $\lambda$} & 
\colhead{STIS Flux\tablenotemark{*}} & \colhead{IUE Flux\tablenotemark{*}} & 
\colhead{FOS Flux\tablenotemark{*}}\\
\colhead{} & \colhead{(\AA)} & 
\multicolumn{3}{c}{(10$^{-3}\,$photons\,\,cm$^{-2}\,$s$^{-1}$)}
}\startdata
{\bf ${}^{5}$D$^{0}\to\,{}^{3}$P} & {\bf 1479} & {\bf 0.51 $\pm$ 0.09} & \nodata & \nodata \\
\nodata & 1473.0 & 0.376 $\pm$ 0.083 & \nodata & \nodata  \\
\nodata & 1481.7 & 0.107 $\pm$ 0.027 & \nodata & \nodata  \\
\nodata & 1485.6 & 0.026 $\pm$ 0.007 & \nodata & \nodata  \\

\tableline

{\bf ${}^{3}$D$^{0}\to\,{}^{3}$P} & {\bf 1479} & {\bf 0.76 $\pm$ 0.11} & \nodata & \nodata \\
\nodata & 1474.1 & 0.394 $\pm$ 0.059 & \nodata & \nodata  \\
\nodata & 1483.1 & 0.289 $\pm$ 0.087 & \nodata & \nodata  \\
\nodata & 1487.2 & 0.075 $\pm$ 0.019 & \nodata & \nodata  \\

\tableline

{\bf ${}^{3}$S$^{0}\to\,{}^{3}$P} & {\bf 1814} & \nodata & {\bf 5.4 $\pm$ 0.9} & {\bf 1.1 $\pm$ 0.2}  \\
\nodata & 1807.3 & \nodata & 1.7 $\pm$ 0.4 & 0.45 $\pm$ 0.08 \\
\nodata & 1820.3 & \nodata & 2.8 $\pm$ 0.4 & 0.50 $\pm$ 0.07 \\
\nodata & 1826.2 & \nodata & 0.5 $\pm$ 0.2 & 0.10 $\pm$ 0.03 \\

\tableline

{\bf ${}^{5}$S$^{0}\to\,{}^{3}$P} & {\bf 1900} & \nodata & {\bf 5.1 $\pm$ 0.9} & {\bf 2.0 $\pm$ 0.2}  \\
\nodata & 1900.3 & \nodata & 3.8 $\pm$ 0.5 & 1.48 $\pm$ 0.09 \\
\nodata & 1914.7 & \nodata & 1.3 $\pm$ 0.4 & 0.43 $\pm$ 0.08 \\

\enddata
\tablenotetext{*}{Flux values are from entries in bold in Table 1.}
\end{deluxetable}

\clearpage

\begin{figure}
\epsscale{.80}
\plotone{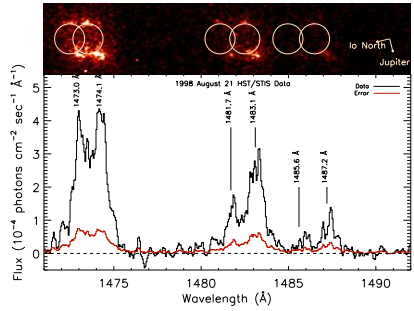}
\caption{Data extracted from the sum of the first two HST/STIS 
G140M exposures in the near vicinity of Io.  The multiplet components of the  
\ion{S}{1} $\lambda$1479 emission are seen.  The brightest emission is located 
at the anti- and sub-jovian equatorial spots (top) with the compass showing 
the direction of Jupiter and Io North.  The flux spectrum corresponding to the 
sub-jovian spot is plotted (bottom) with the data in black and the propagated 
errors in red. }
\end{figure}

\begin{figure}
\epsscale{.80}
\plotone{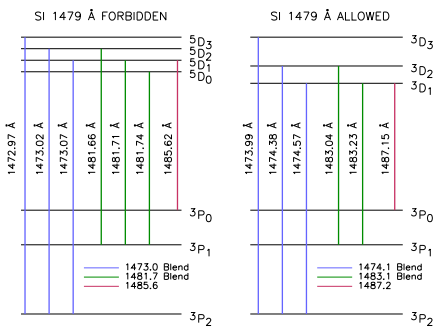}
\caption{Grotrian diagram for the \ion{S}{1} $\lambda$1479 
dipole allowed and forbidden multiplets.  All transition lines are shown.  
The lines indicated as blends are not individually resolved in the STIS data.}
\end{figure}

\begin{figure}
\epsscale{.80}
\plotone{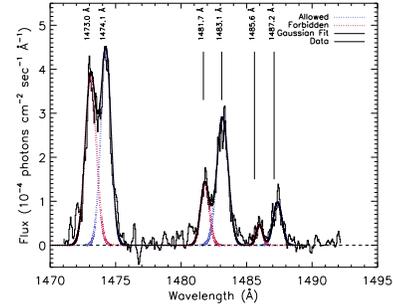}
\caption{Synthetic fit to the flux spectrum of the STIS data 
previously shown in Fig. 1.  The forbidden multiplet fit is in red, the 
allowed in blue, and the total fit in boldface black.  The resolved transition 
lines and blends are indicated.}
\end{figure}

\begin{figure}
\epsscale{.80}
\plotone{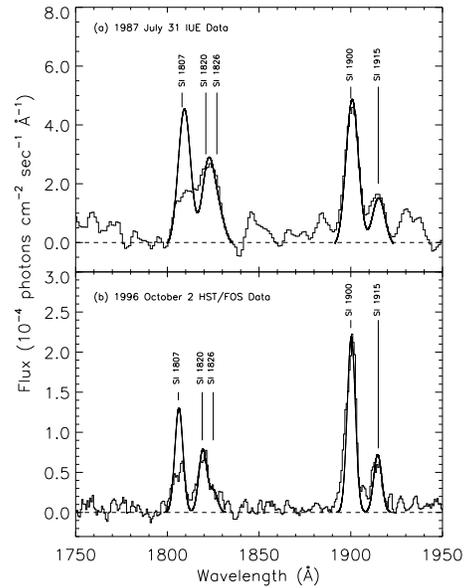}
\caption{Flux spectra of Io's atmospheric emissions in the 
wavelength range 1750-1950 \AA\, taken with (a) IUE and (b) HST/FOS.  
\ion{S}{1} $\lambda$1814 and \ion{S}{1}] $\lambda$1900 emissions are present 
and are overplotted with an optically thin profile of the multiplets in bold.  
The optically thin profile is normalized to the blend of the 1820 \AA\, and 
the 1826 \AA\, line.  This comparison establishes that the 
\ion{S}{1}] $\lambda$1900 transition is optically thin whereas 
\ion{S}{1} $\lambda$1814 is optically thick.}
\end{figure}

\begin{figure}
\epsscale{.80}
\plotone{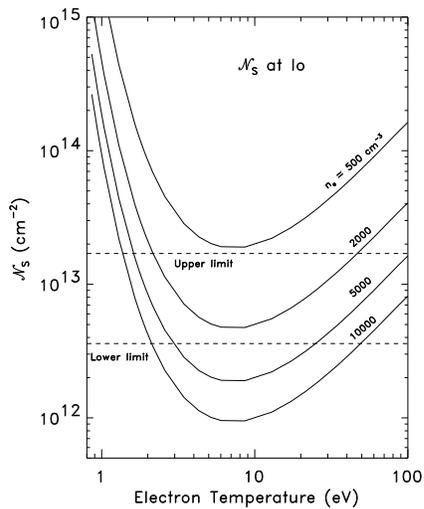}
\caption{Several sulfur column density curves 
corresponding to different values of the electron density are plotted for 
the average measured brightness of \ion{S}{1}] $\lambda$1900 taken from the 
IUE data with Io west of Jupiter.  The upper and lower limits of the sulfur 
column density from our analysis of the STIS \ion{S}{1} $\lambda$1479 data, 
3.6$\times$10$^{12}$\,cm$^{-2}\,<\,{\cal N}_{s}\,<$\,\,1.7$\times$10$^{13}$\,cm$^{-2}$, 
are indicated by the dashed lines.  The canonical $T_e$ and $n_e$ values of 
5 eV and 2000 cm$^{-3}$ give an ${\cal N}_{s}$ of 
5$\times$10$^{12}$\,cm$^{-2}$, consistent with our bounds.}
\end{figure}

\begin{figure}
\epsscale{.80}
\plotone{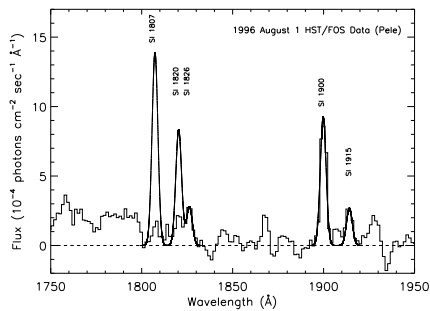}
\caption{Pele volcano flux spectrum from \citet{mcg00}.  A 
synthetic fit for optically thin neutral sulfur emission, used to analyze 
the FOS data presented earlier in this paper, is overplotted on the data 
and is normalized to the 1826 \AA\, line.  This illustrates that the 
\ion{S}{1} $\lambda$1814 multiplet experiences severe optical thickness 
effects while the \ion{S}{1}] $\lambda$1900 doublet does not.  This indicates 
that the sulfur column density is larger over Pele than the equatorial 
spots.}
\end{figure}

\end{document}